\documentclass[]{spie}  

\usepackage{amsmath,amsfonts,amssymb,enumitem}
\usepackage{graphicx}
\usepackage[colorlinks=true, allcolors=blue]{hyperref}
\setlist{nolistsep}

\title{Developing an integrated concept for the E-ELT Multi-Object Spectrograph (MOSAIC): design issues and trade-offs}

\author[a]{Myriam Rodrigues}
\author[b,c]{Gavin Dalton}
\author[d]{Ewan Fitzsimons}
\author[a]{Fanny Chemla}
\author[e]{Tim Morris}
\author[a]{Francois Hammer}
\author[a]{Mathieu Puech}
\author[a]{Christopher Evans}
\author[a]{Pascal Jagourel}

\affil[a]{GEPI, Obs. de Paris, CNRS, Univ. Paris Diderot, 5 Place J. Janssen, 92190 Meudon, France}
\affil[b]{Oxford University, Astrophysics, Denys Wilkinson Building, Keble Road, Oxford OX1 3RH, UK}
\affil[c]{STFC--RALSpace, Harwell Oxford, OX11 0QX, UK}
\affil[d]{UK Astronomy Technology Centre, Royal Observatory Edinburgh, Blackford Hill, Edinburgh, UK}
\affil[e]{Durham University, Department of Physics, South Road, Durham, DH1 3LE, UK}

\begin{document} 
\maketitle

\begin{abstract}
We present a discussion of the design issues and trade-offs that have been considered in putting together a new concept for MOSAIC\cite{Hammer14,Hammer16}, the multi-object spectrograph for the E-ELT. MOSAIC aims to address the combined science cases for E-ELT MOS that arose from the earlier studies of the multi-object and multi-adaptive optics instruments (see MOSAIC science requirements in [\cite{Evans16}]). MOSAIC combines the advantages of a highly-multiplexed instrument targeting single-point objects with one which has a more modest multiplex but can spatially resolve a source with high resolution (IFU). These will span across two wavebands: visible and near-infrared. 
\end{abstract}



\section{MOSAIC Concept}
\label{sect:intro}  
{MOSAIC} is the proposed multiple-object spectrograph for the E-ELT [\cite{Hammer14,Hammer16}]. The instrument will have both multiplex and multi-IFU capability and  will utilise the widest possible field of view provided by the telescope. The {MOSAIC} top-level instrument requirements [\cite{Evans16}] were building on the comprehensive White Paper [\cite{WhitePaper}] on the scientific case for multi-object spectroscopy on the European ELT. In this paper we present the preliminary design concept and trade-off based on the actual top-level instrument requirements. Key to the concept are two design principles: firstly having a shared focal-plate with multi-function tiles which can serve as pick-offs for any of the modes and AO functions; and secondly utilising shared-slit spectrographs whereby the spectrograph optics and detectors can be re-used between the highly-multiplexed mode and the IFU mode. 

\section{Instrument focal plane concept}

\subsection{Tiles concept and patrol field positionner}

\begin{figure}
   \begin{center}
   \begin{tabular}{c}
   \includegraphics[height=8cm]{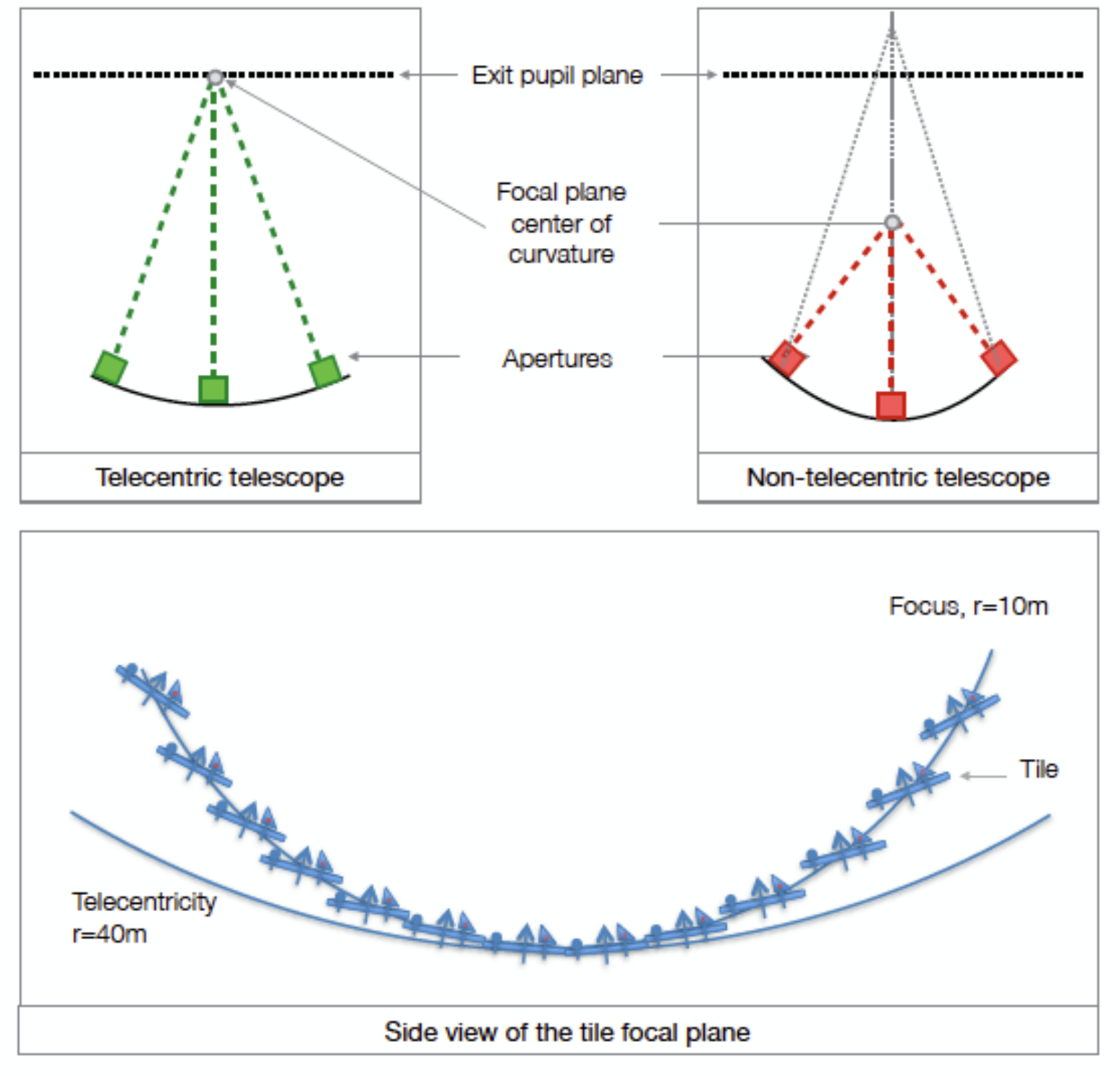}
   \end{tabular}
   \end{center}
   \caption[example] 
   { \label{Fig:non-telecentricity} The \textbf{upper} panels illustrate the ray path in a telecentric telescope (\textbf{left}) and in a non-telecentricity telescope like the E-ELT (\textbf{right}). In a non-telecentric field the light rays do not reach the focal plane perpendicularly but with an incident angle. The bottom panel gives a conceptual view of the tile concept. The focal plane of MOSAIC is covered with hexagonal tiles.  Each tile is set to the mean focus and ray tilt over its patrol field. }
   \end{figure}

\begin{figure}
   \begin{center}
   \begin{tabular}{c}
   \includegraphics[height=10cm]{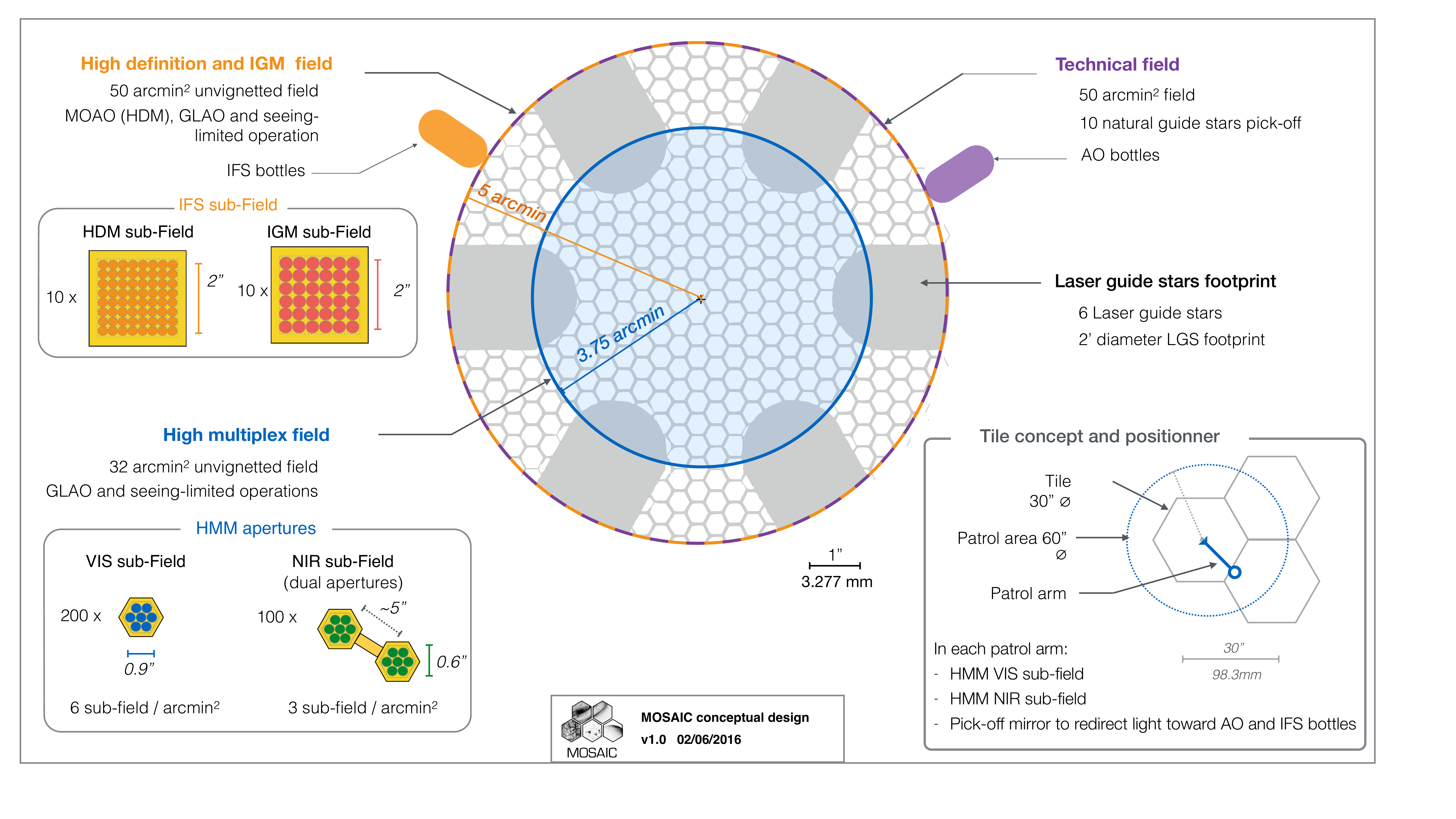}
   \end{tabular}
   \end{center}
   \caption[example] 
   { \label{Fig:Fov_config} 
Conceptual design for the MOSAIC focal plane and the implementation of the 4 observational modes. The diagram illustrating the terminology for the various fields, sub-fields and pick-offs on the MOSAIC focal plane.}
   \end{figure}

The focal plane of the E-ELT is non-telecentric: The light rays do not reach the curved focus plane perpendicularly, but with an incidence angle which depends on the distance to the center of the field, see Figure \ref{Fig:non-telecentricity}. The non-telecentricity of the E-ELT focal plane turns difficult the design of wide field instrument such as MOSAIC. All MOSAIC modes are affected by E-ELT non-telecentricity, but it is particularly harmful for HMM mode which can no longer be self-aligned purely by placing an aperture pickoff on a suitably curved focal plate. Depending on which approach is chosen for the focal plate, the effect of non-telecentricity translates into either a tilt or a defocus effect at the entrance of the subfields. To overcome the non-telecentricity issue, a tile design has been proposed. The concept is to cover the focal plane of MOSAIC with hexagonal tiles and tilt to account for the pupil shift, see figure \ref{Fig:Fov_config}. Each tile is set to the mean focus and ray tilt over its patrol field.  A locally controlled positioner in each tile allocates the HMM apertures and HDM pick-off mirror. This design concept has the advantage to: (1) resolve the non-telecentricity issue locally in each tile; (2) minimise the configuration time and hence removing the need for a carousel. Other positioners have been investigated but they all face-on major drawbacks on non-telecentricity correction and nodding capabilities (see section x). \\

The 4 observational modes aimed for MOSAIC will share this tiled focal plane. Figure \ref{Fig:Fov_config} shows a conceptual design for the MOSAIC focal plane and the implementation of the 4 observational modes. The two highly multiplexed modes (HMM) will operate in seeing limited or ground layer adaptive optics (GLAO) conditions with the following specifications:

\begin{itemize}
\item HMM-VIS: 200 sub-fields within a 3.75 arcmin radius field. Each sub-field consists in bundles of several microlens + fibres.
 \item HMM-NIR: 100 sub-fields consisting in dual apertures for optimal sky-subtraction. 
\end{itemize}

The two integrated field spectroscopy modes HMM and IGM will operate with the following specifications:
\begin{itemize}
 \item HDM: High definition mode, operating with multi object adaptive optics (MOAO) in the near-IR. A pick-off mirror in the focal plane directs light via an MOAO adaptive system (receiver) and fibre bundle to the spectrograph
\item  IGMM: Light bucket IFS operating in seeing limited condition. A pick-off mirror redirect the light via a path compensator and fibre bundle to the spectrograph.
 \end{itemize}
 
\subsection{ High multiplex mode trade-off: allocation efficiency, multiplex and scientific field }

From the science cases, the optimal multiplex in HMM is 200 sub-field within a field of 40 arcmin$^2$[\cite{Evans16}]. From a technical view, the tiled focal plane ties the HMM multiplex to the instrument field of view and tile size. Therefore a trade-off has  been performed between instrument field, the size of the tiles, that set the density of pick-off, and the positioner patrol area. The aim is to maximise the instrument field and maintain the multiplex to 200 sub-fields in the visible, while keeping the allocation of science targets efficient and residuals from non-telecentricity correction low at the edges of the patrol field.

\begin{figure}
   \begin{center}
   \begin{tabular}{c}
   \includegraphics[height=7cm]{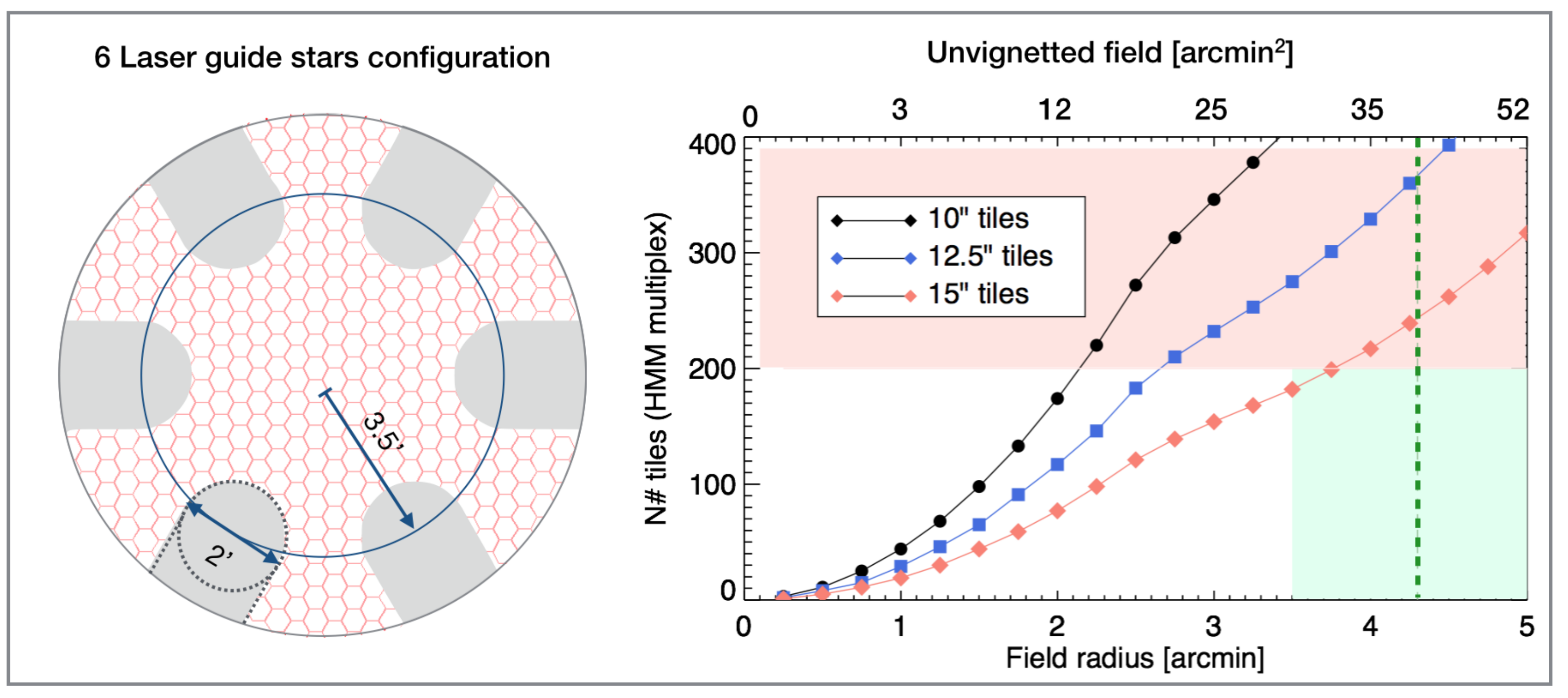}
   \end{tabular}
   \end{center}
   \caption[example] 
   { \label{Fig:Fov_multiplex} \textbf{Left panel}. MOSAIC field for a configuration of 6 laser guide star situated in a circle 3.5$'$ from the center of field. Each LGS asterism make a vignetting of 2' diameter plus. \textbf{Right panel}. Number of tiles as a function of the field radius for three tile radius: 10$"$ (black circles), 12.5$"$ (blue squares), 15$"$ (red diamonds). The green dashed vertical line indicates the top level requirement for the HMM field-of-view 40arcmin2, and the green area a $\pm$10 arcmin$^2$ tolerance. The red region indicates values of multiplex above $\sim$200 (top level requirement).   }
   \end{figure}

In a first approach, we have assumed that each tile will host a single patrol arm positioner. This simplifies the positioner/tile manufacturing and minimises the cost. In this case, the maximum multiplex in HMM modes is given by the number of tiles. The right panel of Figure \ref{Fig:Fov_multiplex} gives the number of tiles (or maximum multiplex) as a function of the field radius for three tiles radius - 10$"$, 12.5$"$, 15$"$ - in a 6 laser guide stars configuration (left panel).  To reach the top level requirement of 40 arcmin$^2$ instrument field and 200 multiplex in HMM-VIS, the option with 30$"$ diameter tiles should be favour.  \\
 
The efficiency of target allocation has been tested for 3 HMM science cases and several combination of positioner density ( tile diameter) and patrol area. The HMM apertures were automatically allocated to the targets using a stable marriage algorithm. The success of the fiber allocation for each SC has been simulated in real fields, see the description of the input catalogues in table \ref{tab:Allocation}. Each combination of \textit{(tile diameter, patrol area)} has been simulated 100 times. At each iteration, the center of the MOSAIC field has been randomly move inside the input catalogue field. The success rate is defined as the number of allocated fibers over the number of available targets in the field-of-view. 
SC1 and SC4/a are observational cases where the density of targets in the field of view is smaller than the density of HMM subfields. In these cases the configuration with a patrol area ranging the center of the adjacent tiles are optimal, with success rate close to 100\% independently of the diameter of the tile (15$"$ to 30$"$). The gain on target allocation to increase the patrol beyond the center of the next tiles is negligible. SC4/b is an observational case where the number of targets in the fov is larger than the multiplex. The success of the target allocation depends on the density of positioners. The SC3 corresponds to a case where the density of potential targets is close to the density of HMM subfields. In this case, the optimal configuration would be tiles of 15$"$ diameter and with a patrol area ranging the center of the adjacent tiles. This configuration gives a success rate of 90\% but need a large number of tiles/fiber, nearly 360. A configuration 30$"$ diameter tiles and a patrol radius ranging the center of adjacent tiles still gives acceptable success rate about 70\%. 

\begin{table}   
\label{tab:Allocation}
\begin{center}       
\begin{tabular}{|l|l|l|l|} 
\hline
SC& SC3: High-z dwarves	& SC1: First galaxies&SC4: Extragalactic pop.\\
\hline
Input & Photometric catalogue of&Catalogues from Bouwens of z=7 & a)Red supergiants cat.; \\
 & Dahlen 2010 & 2 fields simulated : & b)NGC55 field \\
 & - Dwarf galaxies: & -64.5 arcmin2&  \\
 & $log M_*< 9.10^9 M_ \odot$ &-34.2 arcmin$^2$&  \\
 & -[OIII] detected in J-band: & &  \\
 & z=[0.88,1.16] &&  \\
\hline 
Remarks &Clustered distribution& Target density $\sim$23/fov & a)Target density $\sim$35/fov\\
 &Target density $\sim$230/fov&  & clustered distribution\\
 & & &b)Target density $\sim$32000/fov\\
 & & &gradiant distribution\\
\hline 
\end{tabular}
\end{center}
\end{table}

\begin{figure}
   \begin{center}
   \begin{tabular}{c}
   \includegraphics[height=7cm]{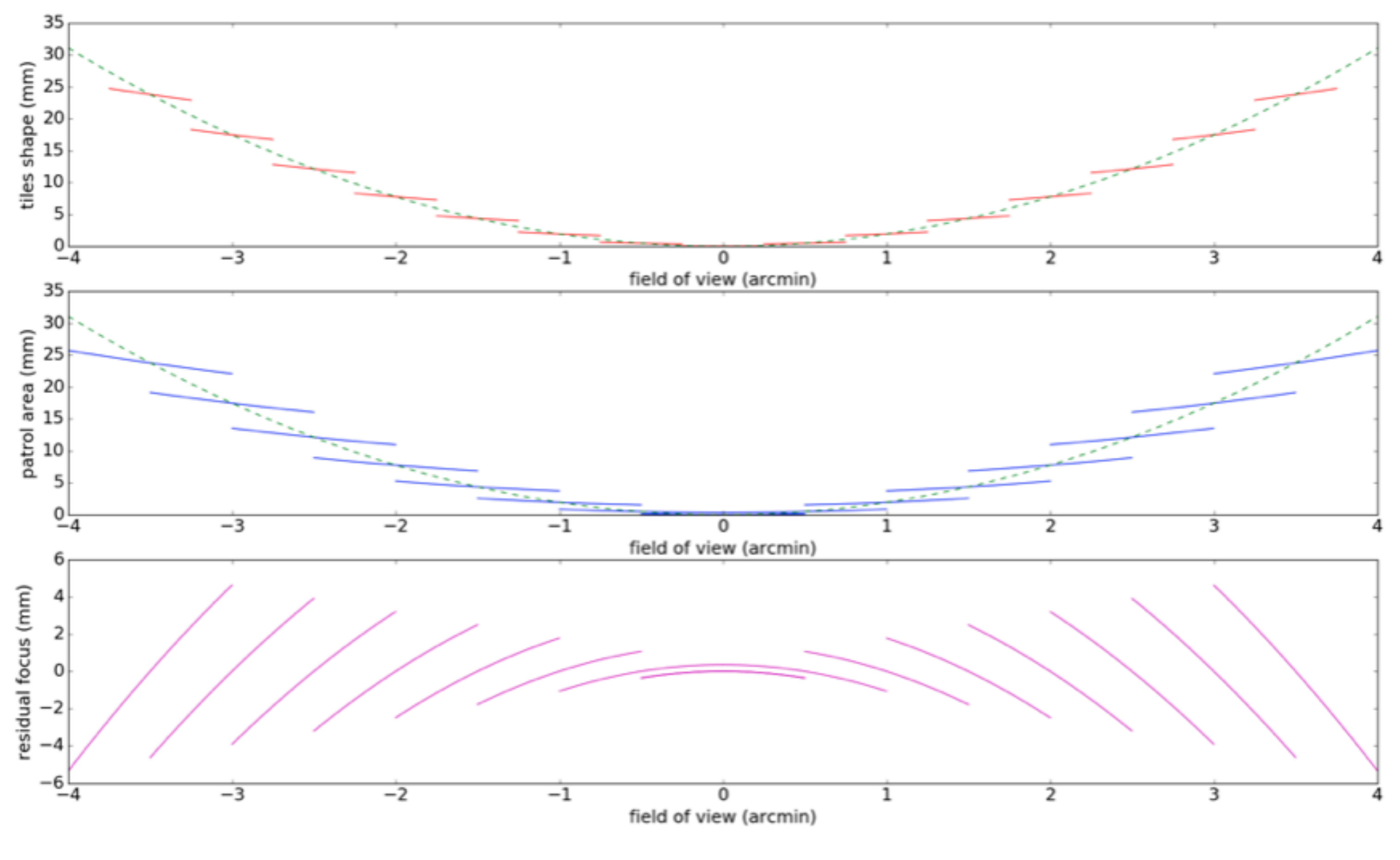}
   \end{tabular}
   \end{center}
   \caption[example] 
   { \label{Fig:Residuals} \textbf{Upper panel}:  The shape of the tiled focal plane (red curve). The green dashed lines represent the field curvature, where the defocus equals zero. \textbf{Middle panel}: The patrol area associated to each tile (blue curve), which is twice the size of a tile.  \textbf{Bottom panel}: The residual defocus on each tile (magenta curve). }
   \end{figure}

Finally, we have investigated the amount of defocus from non-telecentricity residuals as a function of the tile size. Each tile is oriented towards the centre of the pupil, and offset in order to compensate for the defocus. However, a residual defocus remains, which increases with the size of the tile.  A study has been performed in order to evaluate this residual defocus in the case of 30" tiles  with  60" patrol area. The calculation consists in considering a focal plate which radius of curvature is 37.2 m in order to be compliant with the E-ELT exit pupil position, ensuring self-aligned targets, fractioning this focal plate into tiles, and comparing the focus position of each point of the patrol area to the perfect focus, located on a 9.9 m sphere. The results are summarise in Figure \ref{Fig:Residuals}. The defocus has no impact on the pupil conjugation (location and focus) onto the fibrer core, since the exit pupil of the E-ELT can be considered at infinity with respect to the microlens focal length.  No resolution is required inside the HMM sub-field, thus having the field microlens out of focus is not an issue either.  The only effect of defocus could be a loss of flux at the edge of the sub-field due to the enlargement of the seeing or GLAO corrected PSF. The residual focus curve (magenta) shows a maximum defocus for the most off-centre tiles of 5 mm that converts at F/17.48 into less than 0.3 mm PSF enlargement. Considering the E-ELT plate scale, this is less than 0.1"  that could very likely be accommodated passively, removing the need for extra focus compensation on each tile. Confirmation of this, however, is pending the AO analysis on seeing-limited and GLAO-based operation.

\subsection{ Technical field and AO performance}

As part of the adaptive optics work package, a trade-off analysis have been started that includes: instrument field, required ensquared energy, implemented AO modes, number and magnitude of Natural Guide Stars, and number of LGS. A complete description of this trade-off analysis can be found in [\cite{Morris16}].  As part of this a comprehensive set of baseline parameters for AO simulations has been formulated to ensure consistency and interoperability between the various simulator environments being used (at Durham, LESIA, ONERA and LAM). In addition, four main architecture trades were identified which will be prioritised due to their impact on the rest of the system, namely: the number of LGS to be use (0, 4 and 6 will be investigated), whether or not to use MEMS deformable mirror, whether the laser guide stars should track the pupil or the sky and an assessment of the adaptive optics performance with ground layer adaptive optics only.

\section{Sensitivity requirements and sky subtraction}   

MOSAIC will observe extremely faint sources, up to J/HAB $\sim30\,$mag in emission, and up to J/HAB $\sim27\,$mag for continuum and absorption line features in the near-infrared window (e.g. SC1 and SC3 science case). The detection and spectroscopic follow-up of these faint sources will require an accurate and precise sky subtraction process. Accurate sky subtraction is particularly challenging in the near-infrared, where the sky signal is dominated by fluctuating bright sky lines from the airglow. The spectral features from faint sources will be typically observed between these bright OH sky lines. However, the near-infrared (NIR) sky continuum background is still hundreds to a thousand times brighter than the sources to be detected, about J/HAB $\sim19-19.5\,$mag in dark sky conditions (Sullivan \& Simcoe, 2012). For the future detection of such faint sources, the sky continuum in the NIR will need to be subtracted with accuracies at a level of a few tenths of a percent at least.

This first analysis mainly focus on the sky subtraction in the HMM mode. Contrary to HDM, the sky cannot be sampled in the immediate vicinity of the target in HMM. To achieve high accuracy sky subtraction in the HMM, the sky need to be sampled to a distance from the science target inferior to the typical spatial variation of the sky continuum. Yang et al. 2012, Puech et al. 2012 have shown that sky continuum background exhibits spatial variations over scales from $\sim$ 10 to $\sim$ 150 $"$, with total amplitudes below 0.5\% of the mean sky background. At scales of $\sim$10$"$, the amplitude of the variations is found to be $\sim$0.3-0.7\%. Observationally, this small scale fluctuation of the sky background implies that the sky should be sampled less than 5$"$ from the object, and translates into the requirement on a strict upper limit on the minimal distance between fiber. 

Two observational strategy has been defined to achieve high precision sky subtraction with MOSAIC in HMM:
\begin{itemize}
\item Nodding. This sky subtraction strategy will be use in the visible for the science cases requiring accurate sky subtraction. The object and the sky are alternatively by a sub-field following a sequence ABBA or ABAB, obtained by nodding either the telescope or the sub-field. 

\item Cross Beam Switching. This sky subtraction strategy will be use in the near-IR.  The sky is sampled simultaneously at $<5"$ from each object by a sky sub-field. Each science target has two sub-field speared by less 5$"$, forming a dual aperture, see NIR sub-field in Figure \ref{Fig:Fov_config}. The object is observed in both fibers following a sequence ABBA or ABAB, obtained by nodding either the telescope or the sub-field.  During the consecutive A-B sequences, a given object is always observed by one of the fibres bundle of the pairs alternately. This method has the advantage to be similar to the nodding along slit and thus is 100\% of the time on the scientific targets and allows a very accurate instrumental response subtraction. This configuration implies to dedicate half of the sub-fields to sample the sky. 
\end{itemize}

There is no show-stopper to implement these two strategies in the tiles design. Compared to previous design (pick-off positioner), the tiles design increases the complexity on the target allocation software (preparation software) and on the positioning algorithm (configuration sequence of the local positioners). However, MOONS/VLT(phase B), which has a similar local positioner design and will use the same sky subtraction strategies, as successfully complete the first stage of development of such software.

\section{Spectrograph preliminary design constraints}

   \begin{figure}
   \begin{center}
   \begin{tabular}{c}
   \includegraphics[height=4cm]{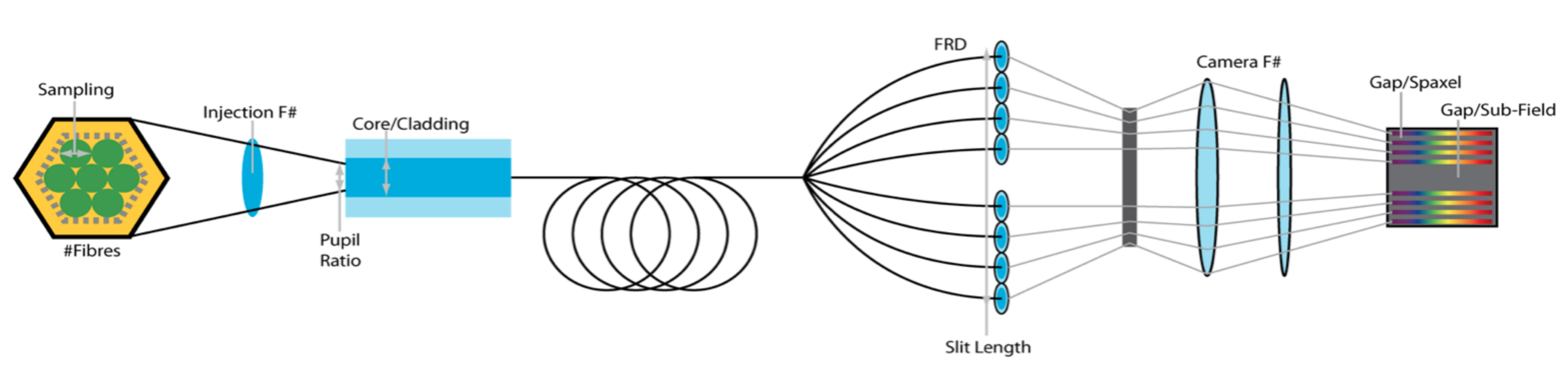}
   \end{tabular}
   \end{center}
   \caption[example] 
   { \label{Fig:Spectrograph_config}
	  Diagram illustrating the HMM-NIR optical path and some of the key parameters which must be optimised.}
   \end{figure} 
   
The plate scale of the E-ELT implies a new constraint for instruments: considering that working at fast F ratio for camera optics is risky, this necessarily implies either more slicing at the field entrance level, or oversampling at the detector level, affecting the multiplex and the spectral coverage. Considering the 0.3" sampling of the visible HMM, an optimal sampling of 2 pixels could only be obtained with a extremely fast camera at around F/0.5 which is not feasible.  

It is complicated further due to the baseline concept of sharing spectrographs - that is utilising a single collimator, grating, camera and detector to serve more than one mode of operation. This concept is considered to be crucial in order to achieve a reasonable multiplex in all modes for a reasonable cost. For MOSAIC, the intention is to have a visible spectrograph serving the HMM-VIS and IGM modes and a near infra-red spectrograph serving the HMM-NIR and HDM modes. The NIR channel is a particular challenge due to the significant difference in sizes between the individual spaxels of the two modes (around 0.075$"$ and 0.2$"$).

Two possible approaches are currently being investigated, both of which aim to keep reasonable parameters for the spectrograph camera:
\begin{enumerate}
\item  Keep a reasonable spectrograph camera F ratio between F/1.5 and F/2, and work with an oversampled PSF onto the detector. The number of pixels per element could reach 6 to 7 pixels for HMM. This approach is detector-consuming but relaxes the constraints on the optics of the spectrograph.
\item  Slice the fibre output into three with an image slicer. This has a negative impact on the multiplex per detector (decreased by 40 \%) but it allows proper sampling to be achieved with a reasonable camera. Further, it presents a great advantage especially for the near infra-red spectrograph: the slit widths of HDM and HMM would be roughly equal, greatly simplifying the overhead required in sharing the spectrographs between the two modes. The effective slit is three times smaller than in the approach (1), making a classical optical design for the camera with a reasonable F ratio possible. The slicing feasibility is under investigation.
\end{enumerate}

\subsection{Pupil Sheer}

The allowable offset of the pupil on a fibre end, the \textit{pupil sheer}, is a critical parameter. It is not feasible to inject a pupil that is exactly the same size as the fibre core, as various manufacturing and alignment tolerances, plus telescope stability, will lead to offsets of the pupil from the centre. This can cause variable clipping that will render any flat-field calibration useless. Instead, the pupil must be over or under sized with respect to the fibre core such that the transmission is constant irrespective of the pupil motion, and thus the calibration is stable. 

A budget for the allowable offset of the pupil, taking into account all known sources of pupil offset at present, has been developed. With a substantial technical immaturity margin applied, an over (or under) sizing of the pupil by 20\% will be targeted. Whether an over- or under-size will be used is still to be determined, and will depend on the required throughput (which is less for an oversize pupil), the size of the spectrograph optics (which increase for an undersized pupil), and the manufacturability of the fibre (which decreases for an oversized pupil).

\section{Detectors: the challenge of OH-line saturation}

The saturation of strong sky lines are a well known limitation in near infrared observations. This is especially critical in bands above 1.4$\mu$m, where sky lines are almost 10 times more intense than in Y and J-band. The presence of these strong sky lines have limited typical DIT in the near-IR to 900s on 10m class telescopes. Because of the large on-sky pixel size, MOSAIC observations will be particularly affected by the quick saturation of sky lines. We have investigated the impact of strong sky lines on maximum exposure times possible in MOSAIC and on science operations. 

\subsection{Sky line saturation and the RoN limited regime}

We have calculated boundaries for optimal exposure times in the HMM in several bandwidth: a minimal DIT, under which observations are RoN dominated; a maximal DIT, over which the strongest sky lines in the bandwidth start to saturate. We have also estimated an optimal DIT: (1) In the visible, the optimal DIT was set as the exposure time for which the RoN noise account for 25\% of the variance of the sky background; (2) In the near IR, the optimal DIT was set to 900s, which is the time scale of variation of the sky. 

This analysis is based on sky spectra simulations using current MOSAIC baseline. In this analysis, we have supposed that the MOSAIC design is optimised for the low spectral resolution mode.  Table 1 gives the current specification of MOSAIC spectrographs and detectors for the HMM mode. We have assumed that the two spectrographs use Volume Phase Holographic gratings. A E2V CCD231-84 15$\mu m$ CCD was assumed for the visible spectrograph and a Teledyn H4RG 15$\mu m$ for the near-IR spectrograph. The E2V CCD231-84 15$\mu m$ specification were gathered from the specification of the MUSE/VLT detector (ref). The Teledyn H4RG 15$\mu m$ specification were assumed to be similar to those of KMOS detector H2RG 18$\mu m$ [ref]

\subsubsection*{Minimal DIT}
Observations should preferentially be carried out in background-limited regime, in which the noise budget is dominated by the Poisson noise of the sky continuum (interlines). In the case of the HMM mode, the noise budget has been calculated within a spaxel (fiber). Figure 6 gives for a spaxel (fiber) the total RMS noise ($N_{spaxel}$) as a function of the sky continuum variance ($N_{bg}^{spaxel}$), for the HMM mode R=5000 in r-band (left panel) and HMM mode R=5000 in H-band (right panel). 

\begin{equation}
N_{bg}^{spaxel} = I_{bg}^{pixel} \times n^2_{pixel} 
\end{equation}

\begin{equation}
N^{spaxel} = \sqrt{ N_{bg}^{spaxel} + n^2_{pixel}RoN^2 + n^2_{pixel} \times dark \times t_{exp}}
\end{equation}

where $I_{bg}^{pixel}$ is the mean photons counts on the sky continuum over the observed bandwidth, RoN is the read-out-noise, dark is the dark current, text is the exposure time, and $n^2_{pixel}$ is the number of pixels over which a spaxel is imaged on the detector. 
Observations are RoN-dominated when $n^2_{pixel}RoN^2 > N_{bg}^{spaxel}$. The minimum DIT for background-limited regime is thus given by:

\begin{equation}
DIT_{min} = \frac{RoN^2}{C_{bg}^{pixel}} 
\end{equation}

where $C_{bg}^{pixel}$ is the count rate (ph/s) in the sky continuum $C_{bg}^{pixel} = I_{bg}^{pixel}/t_{exp}$

\subsubsection*{Maximal DIT}
The maximal DIT was calculated from the counts/s of the strongest sky line ($C_{skyline}^{pixel}$) in a single pixel and the saturation limit of the detector. Because the intensity of OH sky lines fluctuates by more than 20\%, the maximum DIT has been calculated assuming a margin of 2/3 of the detector saturation threshold.  Maximal DIT is given by:
\begin{equation}
DIT_{max} = \frac{2}{3} \frac{Saturation}{C_{skyline}^{pixel}} 
\end{equation}

\subsubsection*{Results}

Table \ref{fig:Saturation} gives the maximum and minimum DITs, for the two resolution settings of the HMM mode, in r- J- and H-bands. The saturation of strong emission lines is particularly problematic for J- and H-band observations, for which sky lines start to saturate before background-limited observation can be reach. The high spectral resolution mode is more affected because of its lower spectral and spatial sampling. The saturation of bright sky line is also affects in a similar way the HDM mode. The quick saturation of sky lines leads to either a read-noise limited performance, or a very short integration time, and thus to poor observation efficiency. In RoN-limited regime, the penalty in term of signal-to-noise ratio scales with the square root of number of exposures.

   \begin{figure}
   \begin{center}
   \begin{tabular}{c}
   \includegraphics[height=8cm]{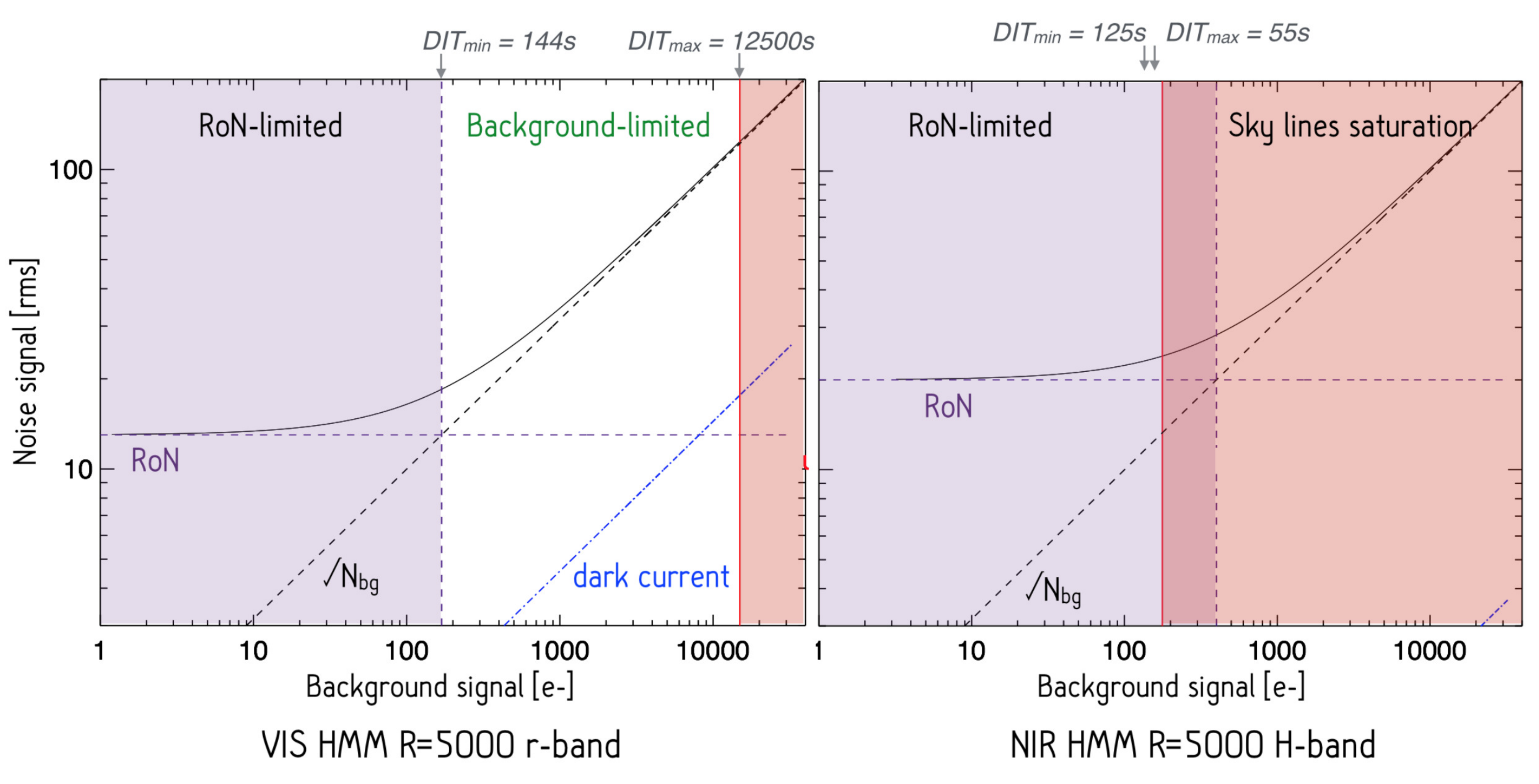}
   \end{tabular}
   \end{center}
   \caption[example] 
   { \label{fig:example} 
Signal-to-Noise regimes in two modes: HMM mode R=5000 in r-band (left) and HMM mode R=15000 in J-band (right). The violet area indicates the regions where the noise is RoN-limited. The red area indicates the region where sky lines are saturated. In the HMM R=5000 in r-band, a background-limited regime can be reach with exposure time above $DIT_{min}$ = 144s. The saturation of sky lines start to be problematic at very high DITs ($DIT_{max}$ > 12500s). In the HMM R=5000 in H-band, background-limited regime cannot be reach with the actual detector and spectrograph configuration. Sky lines are saturated below the minimum DIT to work in a background-limited regime.  }
   \end{figure}

   \begin{figure}
   \begin{center}
   \begin{tabular}{c}
   \includegraphics[height=8cm]{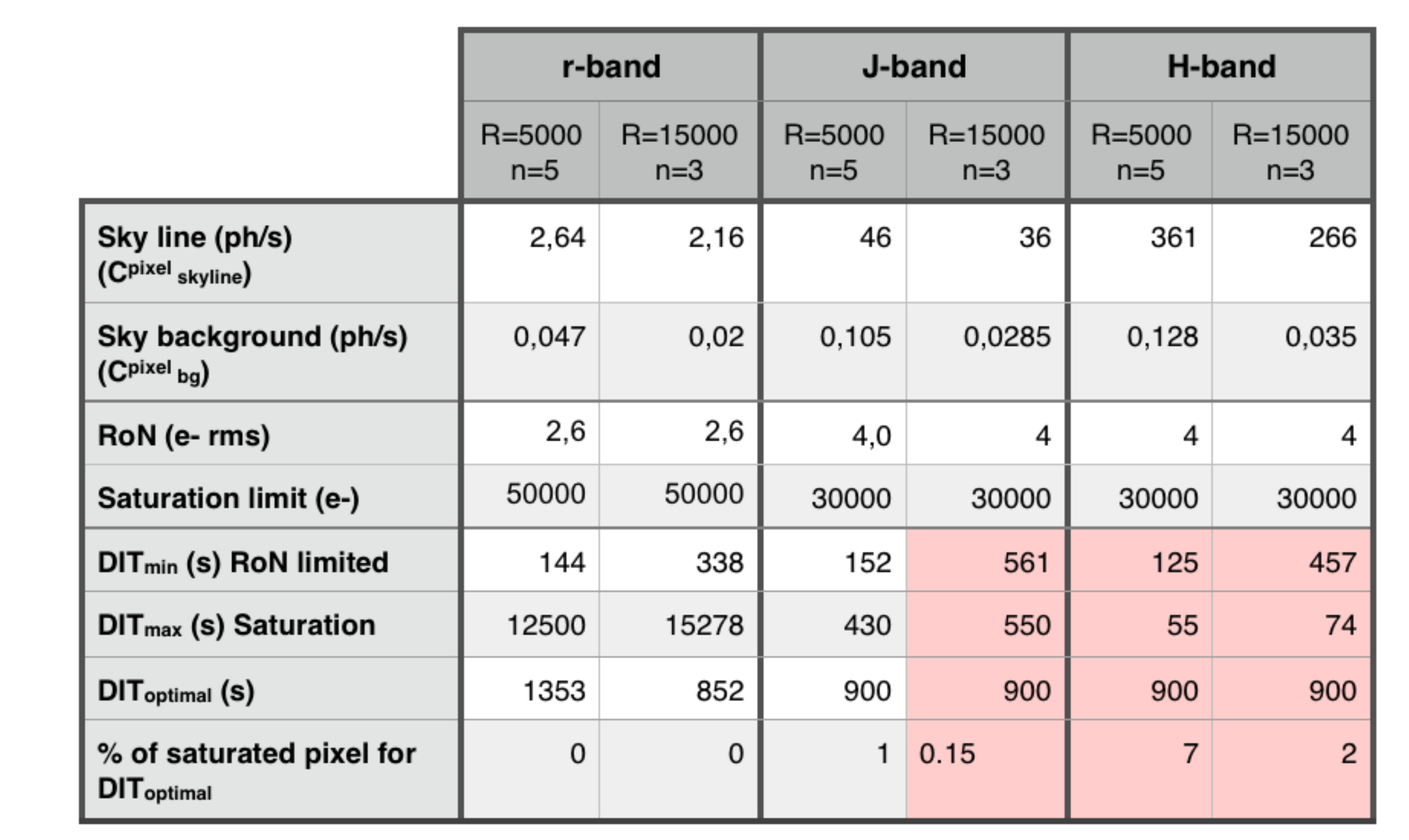}
   \end{tabular}
   \end{center}
   \caption[example] 
   { \label{fig:Saturation} 
 Maximum and minimal DITs for observation in r-,J- and H-band at low and high spectral resolution. The first two line gives the counts of photon per sec of the strongest sky line in the band and the mean background continuum. The cells in red corresponds to modes that are RoN-limited. The last line gives the fraction of pixels saturated assuming optimal DIT observation for the other modes and photometric bands. The fraction of saturated pixels were computed for a full photometric band, and does not take into account the real bandwidth of each mode. }
   \end{figure}

\subsection{Impact on spectrograph design and detector} 

Different combinations of subfield and spectrograph properties (e.g. spaxel diameter,  spectral sampling) have been investigated. It results from this analysis, that the issue of sky line saturation vs RoN-limited observation cannot be resolved by changes on the spectrograph design. During the phase A, solutions will be investigated such as OH-suppressor system \cite{Ellis2012} and skyline masks. The most promising solution is a read mode of CMOS detector which reset the pixels in specific windows while integrating \cite{Bezawada16}. This read mode would permit to integrate during long exposure time (to reach background limited observations), while reseting the saturated pixels several times during the exposure. The fraction of saturated pixels that would need to be removed, in order to reach optimal DIT in each band ($t_{exp} = DIT_{opt}$=900s)., are up to 7\% along the spectral direction (per line in the detector). To minimise the number of rectangular window, the spectrograph concept should minimise the distortion of spectra trace.

\bibliography{report} 

\begin{thebibliography}{1}

\bibitem{Hammer14}
Hammer, F., Barbuy, B., Cuby, J.~G., Kaper, L., Morris, S., Evans, C.~J.,
  Jagourel, P., Dalton, G., Rees, P., Puech, M., Rodrigues, M., Pearson, D.,
  and Disseau, K., ``{MOSAIC} at the {E-ELT}: A multi-object spectrograph for
  astrophysics, {IGM} and cosmology,'' {\em Proc. SPIE}~{\bf 9147},
  914727--914727--7 (2014).

\bibitem{Hammer16}
Hammer, F., Morris, S., Kaper, L., Barbuy, B., Cuby, J.~G., Roth, M., Jagourel,
  P., Evans, C.~J., Puech, M., Fitzsimons, E., Dalton, G., and Rodrigues, ``The
  e-elt multi-object spectrograph: latest news from {MOSAIC},'' {\em Proc.
  SPIE}~{\bf 9908}(9908-78) (2016).

\bibitem{Evans16}
Evans, C., Puech, M., Rodrigues, M., Barbuy, B., Cuby, J.-G., Dalton~G., F.~E.,
  Hammer, F., Jagourel, P., Kaper, L., L., M.~S., Morris, T.~J., and the MOSAIC
  Science~Team, ``Science requirements and trade-offs for the {MOSAIC}
  instrument for the european elt,'' {\em Proc. SPIE}~{\bf 9908}(908-353)
  (2016).

\bibitem{WhitePaper}
Evans, C., Puech, M., and Team, M.~S., ``{The Science Case for Multi-Object
  Spectroscopy on the European ELT},'' {\em ArXiv e-prints}  (Jan. 2015).

\bibitem{Morris16}
Morris, T., Basden, A., Buey, T., Chemla, F., Conan, J., Fitzsimons, E., Fusco,
  T., Gendron, E., Hammer, F., Jagourel, P., Morel, C., Myers, R., Neichel, B.,
  Petit, C., Rodrigues, M., and G., R., ``Adaptive optics for {MOSAIC}: design
  and performance of the wide(st)-field ao system for the e-elt,'' {\em Proc.
  SPIE}~{\bf 9909} (2016).

\bibitem{Ellis2012}
Ellis, S.~C., Bland-Hawthorn, J., Lawrence, J., Horton, A.~J., Trinh, C.,
  Leon-Saval, S.~G., Shortridge, K., Bryant, J., Case, S., Colless, M., Couch,
  W., Freeman, K., Gers, L., Glazebrook, K., Haynes, R., Lee, S.,
  L{\"o}hmannsr{\"o}ben, H.-G., O'Byrne, J., Miziarski, S., Roth, M., Schmidt,
  B., Tinney, C.~G., and Zheng, J., ``Suppression of the near-infrared oh
  night-sky lines with fibre bragg gratings - first results,'' {\em mnras}~{\bf
  425},  1682--1695 (Sept. 2012).

\bibitem{Bezawada16}
Bezawada, N. and Ives, D., ``High-speed multiple window readout of hawaii-1rg
  detector for a radial velocity experiment,'' {\em Proc. SPIE}~{\bf 6276},
  914727--914727--7 (2006).

\end{thebibliography}
\bibliographystyle{spiebib}

\end{document}